\documentclass[aps,pra,superscriptaddress,twocolumn]{revtex4}

\usepackage{bm}
\usepackage{graphicx}
\usepackage{amsmath}

\begin{document}

\title{When finite-size corrections vanish:\\
The $S=1/2$ XXZ model and the Razumov-Stroganov state}

\author{Leonardo Banchi}
\affiliation{Dipartimento di Fisica, Universit\`a di Firenze,
    Via G. Sansone 1, I-50019 Sesto Fiorentino (FI), Italy;}
\affiliation{Istituto Nazionale di Fisica Nucleare, Sezione di Firenze,
    Via G. Sansone 1, I-50019 Sesto Fiorentino (FI), Italy;}
\author{Filippo Colomo}
\affiliation{Istituto Nazionale di Fisica Nucleare, Sezione di Firenze,
    Via G. Sansone 1, I-50019 Sesto Fiorentino (FI), Italy;}
\author{Paola Verrucchi}
\affiliation{Centro di Ricerca e Sviluppo SMC
 dell'Istituto Nazionale di Fisica della Materia -- CNR,
 Sezione di Firenze, via G.~Sansone 1,
 I-50019 Sesto Fiorentino, Italy.}
\affiliation{Dipartimento di Fisica, Universit\`a di Firenze,
    Via G. Sansone 1, I-50019 Sesto Fiorentino (FI), Italy;}
\affiliation{Istituto Nazionale di Fisica Nucleare, Sezione di Firenze,
    Via G. Sansone 1, I-50019 Sesto Fiorentino (FI), Italy;}


\begin{abstract} 

We study the one-dimensional $S=1/2$ XXZ model on a finite lattice at zero 
temperature, varying the exchange anisotropy $\Delta$ and the number of 
sites $N$ of the lattice. Special emphasis is given to the model with 
$\Delta=1/2$ and $N$ odd, whose ground state, the so-called 
Razumov-Stroganov state, has a peculiar structure and no finite-size 
corrections to the energy per site. We find that such model corresponds to 
a special point on the $\Delta$-axis which separates the region where 
adding spin-pairs increases the energy per site from that where the longer 
the chain the lower the energy. Entanglement properties do not hold 
surprises for $\Delta=1/2$ and $N$ odd. Finite-size corrections to the 
energy per site non trivially vanish also in the ferromagnetic $\Delta\to 
-1^+$ isotropic limit, which is consequently addressed; in this case, 
peculiar features of some entanglement properties, due to the finite 
length of the chain and related with the change in the symmetry of the 
Hamiltonian, are evidenced and discussed. In both the above models the 
absence of finite-size corrections to the energy per site is related to a 
peculiar structure of the ground state, which has permitted us to provide 
new exact analytic expressions for some correlation functions.

\end{abstract}

\maketitle


\section{Introduction}
Low dimensional magnetic systems have been acknowledged as intriguing 
physical
systems for decades and still attract much interest, both from the
theoretical and the experimental point of view.  Many reasons justify such
interest, and one more has been recently added, namely the
possibility to use $S=1/2$ spin models as tools for studying problems 
related to quantum information theory and quantum computation 
\cite{DiVincenzoEtal00,MeierLL03,AAVVnato06}.
Amongst one-dimensional systems, a preminent role is played by the 
Heisenberg Hamiltonian
\begin{equation}
{\cal{H}}=\sum_i \left(J_xS_i^xS^x_{i+1} + J_yS_i^yS_{i+1}^y + 
J_zS_i^zS_{i+1}^z\right)\,,
\label{e.HXYZ}
\end{equation}
where $i$ runs over the sites of a chain, and $S_i$ are angular 
momentum operators satisfying 
$[S^\alpha_i,S^\beta_j]=i\delta_{ij}\varepsilon^{\alpha\beta\gamma}S^\gamma_i$
 ($\alpha,\beta,\gamma=x,y,z$).
Models described by the Hamiltonian (\ref{e.HXYZ})
constitute a class which is characterized, even at zero temperature, 
by the possible occurrence of peculiar phenomena, such as quantum phase 
transitions \cite{Sachdev99}, saturation \cite{YangY66,KatsuraS70,Minami04}, or factorization 
\cite{KTM-82,RoscildeEtal04}.  Whether a specific model 
displays one such
phenomenon depends on the details of the exchange interaction and, in
case, on the value of an external magnetic field. The values of the 
exchange
parameters $J_x,J_y$ and $J_z$ define each model, 
which in fact may get its name after such values. 

In this paper we will refer to the $S=1/2$ XXZ chain with a finite
number of sites $N$ and periodic
boundary conditions, whose Hamiltonian can be written in the 
dimensionless form
\begin{equation}
{\cal{H}}=\sum_{i=1}^N \left[-(\sigma_i^x\sigma_{i+1}^x+\sigma_i^y\sigma_{i+1}^y)+\Delta
\sigma_i^z\sigma_{i+1}^z\right]\,,
\label{e.HXXZ}
\end{equation}
where $\sigma_i^\alpha$ are the Pauli matrices for the spin sitting at 
site $i$. The above expression is obtained from Eq.~(\ref{e.HXYZ}) by 
setting $J_x=J_y=-4$ and $J_z=4 \Delta$. The choice of the minus sign in 
front of the exchange interaction on the $xy$-plane implies no loss of 
generality in the thermodynamic limit, due to 
the possibility of changing such sign at will 
via a unitary transformation. However, when finite 
chains are considered, special care is due to this aspect, as explained in 
Section \ref{s.r-s_model}.

The behaviour of the model depends on 
the value of the anisotropy parameter:
isotropic ferro- ($\Delta=-1$) or antiferromagnetic 
($\Delta=1$); Ising-like ($|\Delta|>1$); critical 
($|\Delta|<1$). Within each interval, the actual value of $\Delta$ is not of 
particular interest, at least as far as the general phenomenology 
is concerned. However, in the case of the XXZ model Eq.~(\ref{e.HXXZ}) 
there exists 
an exception to this statement: In fact, for $\Delta=1/2$,
periodic boundary conditions, and a finite and odd number of sites, the 
ground state of the model, often referred to as the Razumov-Stroganov 
state, shows very peculiar features \cite{ABB-88,S-01,RS-01} which 
still stand as an unintelligible occurrence. Such features do not fall within 
the framework of quantum phase transitions, as
the model lies well inside the $|\Delta|<1$ region of the XXZ
Hamiltonian; moreover, the matter is relevant only as far as the
number of sites of the chain is finite, so that the difference between odd
and even number of sites stays meaningful.

The Razumov-Stroganov state has been studied by several authors (see 
for example Refs.~\cite{PRdGN-02,FNS-03,DFZJ-05,dG-05}), with 
different approaches and in many different frameworks, 
but somehow neglecting the fact that the corresponding
model is not isolated in the phase diagram of the XXZ model.

In this paper, we tackle the problem from a different point of 
view: Given the fact that a very peculiar ground state occurs only for 
$\Delta=1/2$ and odd number of sites, we develop a 
comparative analysis of the behaviour 
of other XXZ models, i.e. models defined by Eq.~\eqref{e.HXXZ} with 
$\Delta\neq 1/2$, and $N$ both even and odd,
looking for clues about what in fact makes the model 
whose ground state is the Razumov-Stroganov state so special.
From a physical perspective, one of the
most interesting feature of the Razumov-Stroganov state is the absence of
finite-size corrections to its energy per site. This property
is shared by the fully separable ground state of the 
finite-length ferromagnetic isotropic ($\Delta=-1$) chain and also by the  
ground state of the $\Delta\to -1^+$ limit, whose structure is however not
as trivial, and deserves special attention.

In the above framework, we have analyzed  the effects of 
the finite length of the chain on the energy, the correlation functions, 
and some entanglement properties.
The analysis presented is based on exact (i.e. available with infinite 
precision) results as far as the Razumov-Stroganov state and the ground state
of the $\Delta\to-1^+$ model are concerned, 
and on numerical data in all the other cases. Numerical diagonalization 
techniques underlie all our outcomes.

The structure of the paper is as follows: In Section \ref{s.r-s_model} we 
introduce the Razumov-Stroganov
state and briefly recall its structure and properties.
Some new exact results for  the two-point correlation functions
of the corresponding chain ($\Delta=1/2$ and $N$ odd) are also presented.
In Section \ref{s.other_models} we consider the energy and the correlation 
functions  
at $T=0$ of several XXZ models, with $|\Delta|<1$ and $N$ both even and odd, 
focusing our attention upon finite-size corrections. 
In Section \ref{s.entanglement} the same type of analysis is proposed 
for some entanglement properties at $T=0$. 
Conclusions are drawn in Section \ref{s.conclusions}.

Some exact analytic expressions for the correlation functions of the
$\Delta=1/2$ and the $\Delta\to-1^+$ models are presented in the two Appendices.


\section{Specificity of the value $\Delta=1/2$}
\label{s.r-s_model}

Let us consider the XXZ Hamiltonian \eqref{e.HXXZ} on a finite chain with 
periodic boundary conditions and $N$ odd, at the particular value $\Delta=1/2$:
\begin{eqnarray}\label{e.HRS}
{}&{}&{\cal{H}}=\sum_{i=1}^N\left[
-(\sigma^x_i\sigma^x_{i+1}
+\sigma^y_i\sigma^y_{i+1})
+\frac{1}{2}\sigma^z_i\sigma^z_{i+1}\right]\,,\nonumber\\
{}&{}&\qquad{\vec{\sigma}}_{N+1}=
{\vec{\sigma}}_{1}\,;\qquad\qquad N\mathrm{\ odd}\,.
\end{eqnarray}
It is worth noticing that, due to the specific conditions on the lattice,  
the negative sign of the transverse coupling cannot be reversed at 
will: Indeed, the rotation of $\pi$ around the $z$-axis of all the spins 
sitting at every other site can be safely performed 
only in the case of $N$ even, or in the thermodynamic limit.

The ground state of the above Hamiltonian is doubly degenerate 
for any $N$ finite and odd, and 
the two degenerate ground states  are eigenstates of 
$\sum_{i=1}^NS_i^z$, with eigenvalue $S^z_{tot}=\pm 1/2$.
The two ground states are obtained from each other reversing the
component along the quantization axis of each spin, and their structure is 
that of the so-called Razumov-Stroganov  state. 
In the following we shall always refer to the  
$S^z_{tot}=+1/2$ case.

The Razumov-Stroganov  state shows several intriguing properties 
\cite{ABB-88,S-01,RS-01}: 
\textit{i)} its energy per site reads exactly $E/N=-\frac{3}{2}$, 
with no finite-size corrections;
\textit{ii)} the coefficients of the ground state
on the {\it standard} basis  (i.e. the basis where all operators $\sigma^z_i$, 
$i=1,\dots N$ are diagonal), are integer multiples of the smallest one;
 \textit{iii)}  some of these integer numbers have a non trivial combinatorial 
interpretation. 

A complete understanding of these features 
is still lacking.
It is worth recalling that the XXZ chain with  $\Delta=1/2$ is known
to have a similarly special ground state in two other cases:
\textit{a}) twisted boundary conditions and $N$ even~\cite{ABB-88,RS-01b};
\textit{b}) open boundary conditions for whatever $N$, even or 
odd~\cite{ABB-88,BdGN-01}.

The fact that, when suitably normalized, the coefficients defining the 
ground state of Eq.~(\ref{e.HRS}) on the standard basis are all integers 
(see point \textit{ii)} above) allows for their \textit{exact} numerical 
evaluation~\cite{B-08,CCN-09}, for $N$ not too large, limited only by 
computing capabilities. Indeed we have computed numerically these 
integer-valued coefficients with infinite precision, on a standard desktop 
computer, for system sizes up to $N=25$.
From the exact knowledge of the ground state the correlation 
functions are readily determined and, as in the present case they
necessarily assume rational values, we have computed them with
infinite precision, again for chain lengths up to $N=25$.
We report in Appendix~\ref{s.AppendixA}, for illustrative purposes, some 
of our results for the  second neighbour longitudinal  two-point correlation
function $\langle\sigma^z_i\sigma^z_{i+2}\rangle_{_N}$.
Here and below $\langle \dots \rangle_{_N}$ denotes the 
expectation value over the ground state of a chain of $N$ spins.
Further results are available upon request~\cite{B-08}.

It appears that the knowledge  of the exact values
of such correlation functions for a finite set of values of $N$, 
together with the fact that these are rational numbers, allows
to determine their analytic expressions as functions of $N$ (obviously, 
only valid for odd $N$).
This is a rather exceptional situation for an interacting critical system.
In particular, from the infinite-precision data mentioned above, 
and the  exact  thermodynamic-limit values computed in 
Refs.~\cite{KMST-05} and 
\cite{SS-07}, we have been able to work out exact analytic
expressions for the $r^{\rm{th}}$-neighbour
longitudinal and transverse two-point correlation functions, 
for $r=1,2,3,4,5$, and arbitrary $N$. The expressions for $r=1$ coincide
with those computed by means of purely analytic methods in Ref.~\cite{S-01}
\begin{equation}\label{e.g+-gzz}
\langle\sigma_i^+\sigma_{i+1}^-\rangle_{_N}=\frac{5}{16}+\frac{3}{16N^2}~,
\quad
\langle\sigma_i^z\sigma_{i+1}^z\rangle_{_N}=-\frac{1}{2}+\frac{3}{2N^2}~,
\end{equation}
while those for $r=2,3,4,5$ were previously unknown. We report them
in Appendix~\ref{s.AppendixA}.

Let us briefly illustrate the method used in deriving such expressions,
taking as an example the case of  the second-neighbour longitudinal  
two-point correlation
function $\langle\sigma^z_i\sigma^z_{i+2}\rangle_{_N}$, whose exact values
for $N=3,5,\dots,15$, are reported in Appendix~\ref{s.AppendixA}.
Subtracting from such values the corresponding exact 
value in the thermodynamic limit (in this case, $7/64$, see 
\cite{KMST-05}),  and factoring into prime numbers the 
resulting denominators, it is easy to infer  for these denominators 
a behaviour, as a function of $N$, of the form $2^6 N^2(N^2-4)$.
The corresponding numerators are also expected to behave  as a  
polynomial in $N$, although of lower order, since the fraction as a whole
should vanish in the thermodynamic limit. The first four values of 
$\langle\sigma^z_i\sigma^z_{i+2}\rangle_{_N}$, for $N=3,5,7,9$, are sufficient
to determine the polynomial in the numerator, thus completely fixing the
expression of $\langle\sigma^z_i\sigma^z_{i+2}\rangle_{_N}$ for arbitrary $N$,
see Eq.~\eqref{e.app.gzzj=2}. The obtained expression reproduces exactly 
the available numerical values of $\langle\sigma^z_i\sigma^z_{i+2}\rangle_{_N}$,
for $N$  larger than $9$. From a strictly
mathematical point of view, the results of such procedure can 
only be conjectural, though strongly supported by the numerical data.

The above  procedure
can be applied to  $r^{\rm{th}}$-neighbour correlation functions, but the 
degree of the involved polynomials increases rapidly. For generic values of
$r$, the denominators appears to behave as 
$N^{2[(r+1)/2]}\prod_{i=1}^{[r/2]}(N^2-4i^2)^{r-2i+1}$. 
When $r=6$, for example,  the denominator is of degree $24$ in $N$, 
and even under the reasonable assumption (supported by all lower-$r$ 
examples)
that the numerator is an even polynomial in $N$,  we need $12$ data to determine it, 
and possibly one more to check the result. 
These data can be taken from the values for $N=7,9,\dots 31$, which however 
lie beyond our computing capabilities. In this respect it is worthwhile to
emphasize that, although the derivation of the results presented above and 
in  Appendix~\ref{s.AppendixA}  rely heavily on computer aided 
evaluations, they are all {\em exact}.


\section{Other models}
\label{s.other_models}

In this Section 
we aim at getting some deeper insight into the physical 
mechanisms possibly related to the peculiar ground state of 
Hamiltonian (\ref{e.HRS});  in particular, we wish to highlight
the specific role played by each of the 
two conditions defining the model, i.e. $\Delta=1/2$ and $N$ finite and 
odd (in the case of periodic boundary conditions); 
to this purpose we develop a 
comparative analysis of the behaviour 
of other XXZ models, i.e. models defined by Eq.~(\ref{e.HXXZ}) with 
$\Delta\neq 1/2$, for different values of $N$. 

We first recall that, despite the most spectacular features being observed 
for $N$ odd, the model (\ref{e.HRS}) has a precise 
specificity 
also in the thermodynamic limit. In fact, as demonstrated by Baxter in 
Ref.~\cite{B-71}, whenever the exchange parameters
 of an infinite Heisenberg chain, Eq.~(\ref{e.HXYZ}), 
satisfy the 
condition 
\begin{equation}
J_xJ_y+J_xJ_z+J_yJ_z=0~
\label{e.BAXTERcondition}
\end{equation}
the energy per site $E/N$ of the 
ground state gets the value 
\begin{equation}
\frac{1}{4}(J_x+J_y+J_z)~;
\label{e.E-special}
\end{equation}
in the case of the XXZ model (\ref{e.HXXZ}), this may
only occur for $\Delta=1/2$, a condition which therefore defines a somehow
{\it special} point, though by no means related with quantum critical
transitions, nor with possible factorization or saturation of the ground 
state.

In Section \ref{s.r-s_model} we have seen that one of the most 
intriguing peculiarities of the model Eq.~(\ref{e.HRS}) is that
 finite-size corrections do not enter the energy expression, which 
necessarily implies $\langle{\cal{H}}\rangle_{_N}/N=-3/2$,
given that  condition   (\ref{e.BAXTERcondition}) is fulfilled.
Due to translation invariance, and given
that ${\cal{H}}=\sum_i{\cal{H}}_{i,i+1}$, the absence of finite-size 
corrections is a property
of the local energy, by this meaning that the expectation value of the
nearest-neighbour  interaction is not affected by the possible 
addition of
no matter how many spin-pairs (though adding one single spin would
drastically change the whole picture, bridging the model to the
essentially different case of $N$ even); the above feature is seen to 
follow  
from the exact cancellation of the $N$-dependent terms in the
nearest-neighbour correlation functions, when combined as necessary: In
fact, by Eqs.~(\ref{e.g+-gzz}), it is 
\begin{eqnarray}\label{e.H_N} 
\langle{\cal{H}}_{i,i+1}\rangle_{_N}&=&
-4\langle\sigma^+_i\sigma^-_{i+1}\rangle_{_N}+
\frac{1}{2}\langle\sigma^z_i\sigma^z_{i+1}\rangle_{_N} 
\nonumber \\ 
&=&-\frac{5}{4}-\frac{3}{4N^2}-\frac{1}{4}+\frac{3}{4N^2}=-\frac{3}{2}~,
\end{eqnarray} 
which shows that, despite the correlation functions being affected by 
finite-size corrections, the interaction energy is not.

\begin{figure}
\includegraphics[width=1\linewidth,height=13truecm]{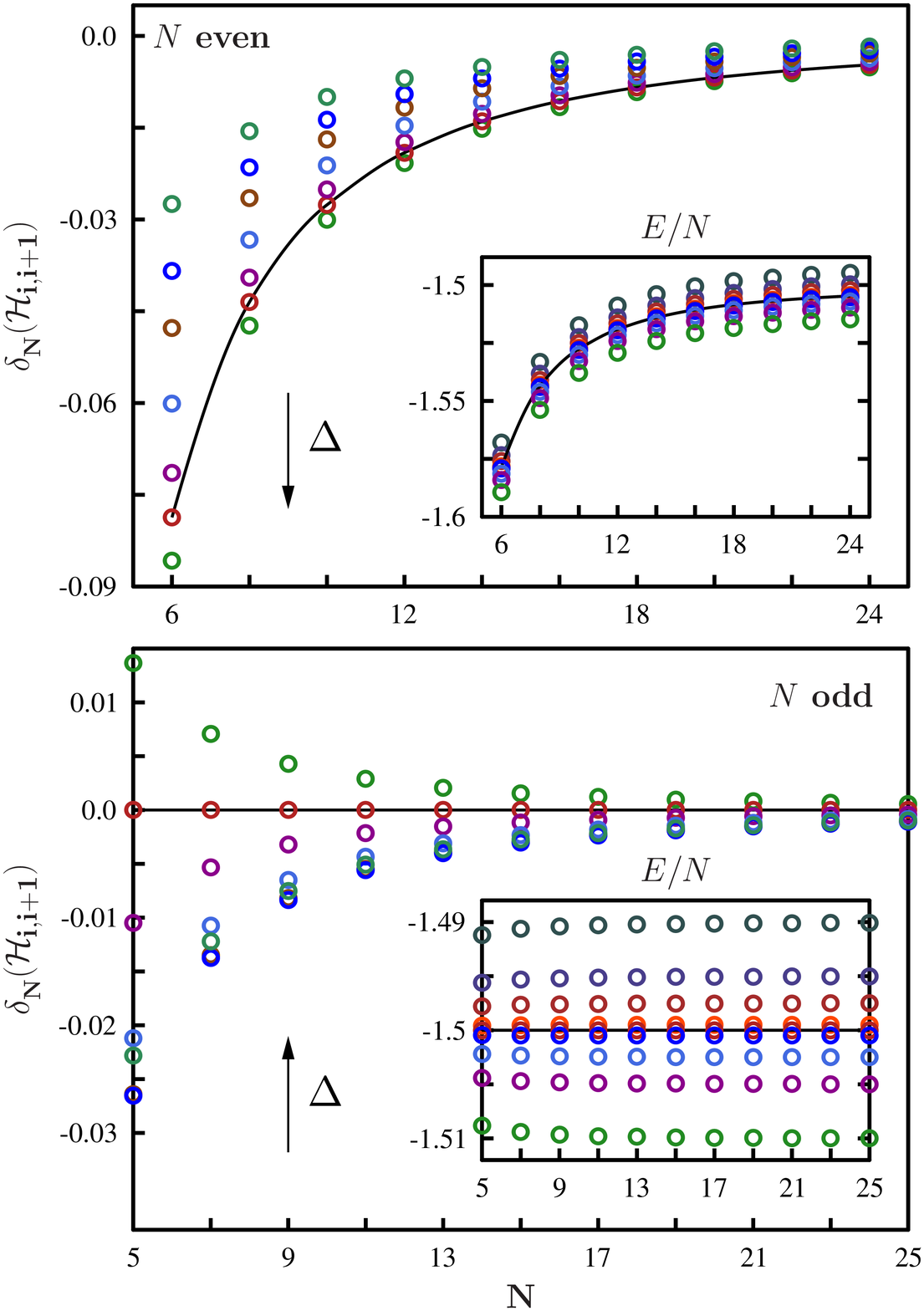}
\caption{$\delta_{_N}({\cal{H}}_{i,i+1})$ versus $N$ for 
$\Delta=$ 0.7, 0.5, 0.3, 0, -0.3, -0.5, -0.7. 
The full line is a guide for the eyes and connects data for $\Delta=1/2$.
Insets: $E/N$  versus $N$ for $\Delta$=
0.48, 0.49, 0.495, 0.499, 0.50, 0.501,  
0.505, 0.51, 0.52.
The anisotropy parameter $\Delta$ grows as indicated by the arrows.
Upper panel: $N$ even; lower panel: $N$ odd.}
\label{f.1}
\end{figure}

Let us now analyze how XXZ models behave for different values of $N$ and 
$\Delta$. For the sake of clarity, the finite-size corrections relative to 
any physical observable ${\cal{O}}$ will be hereafter studied in terms of the 
quantity $\delta_{_N}({\cal{O}})$ defined by
\begin{equation}
\delta_{_N}({\cal{O}})=
\langle{\cal{O}}\rangle_{_N}-
\lim_{M\to\infty}\langle {\cal{O}}\rangle_{_M}~.
\label{e.deltaO}
\end{equation}
We first consider $\delta_{_N}({\cal{H}}_{i,i+1})$ which is shown in 
Fig.~\ref{f.1} for different values of $\Delta$ and as a function of $N$: 
The qualitative difference between even (upper panel) and odd (lower 
panel) number of sites is evident: finite-size corrections 
are negative
for all values of $\Delta$ and $N$, except $\Delta\ge 1/2$ and $N$ odd.

In other terms, while in general the energy per site increases with 
the size of the chain, a special situation occurs for odd $N$ and 
$1/2<\Delta<1$, where adding spin pairs lowers the energy: 
Within the odd-$N$ sector, Hamiltonian \eqref{e.HRS} is 
hence found to correspond to the 
separating case
between these two qualitatively different behaviours of the
finite-size corrections in the energy of the ground state.
The Insets in Fig.~\ref{f.1} show how $E/N$ varies 
with $N$ for various anisotropies in the vicinity of the 
value $\Delta=1/2$: 
It is evident that such value is crossed with continuity, both in the 
odd- and in the even-$N$ sector; this particularly means, in the former 
case, that finite-size corrections smoothly vanish for $\Delta\to 1/2$. 
A simple perturbative analysis for very small $\epsilon=\Delta-1/2$ yields
\begin{equation}
\frac{E}{N}\simeq -\frac{3+\epsilon}{2}+\frac{3\epsilon}{2N^2}~,
\label{e.perturbative}
\end{equation}
in complete agreement with the above scenario.

It is to be noticed that the existence of a unique value of $\Delta$, 
where $\delta_{_N}({\cal{H}}_{i,i+1})$ changes its sign 
{\it 
for all} 
values of $N$ 
odd, is definitely not granted, as shown below in the case of the 
nearest-neighbour correlation functions along the $z$-direction.

\begin{figure}
\includegraphics[width=1\linewidth,height=15truecm]{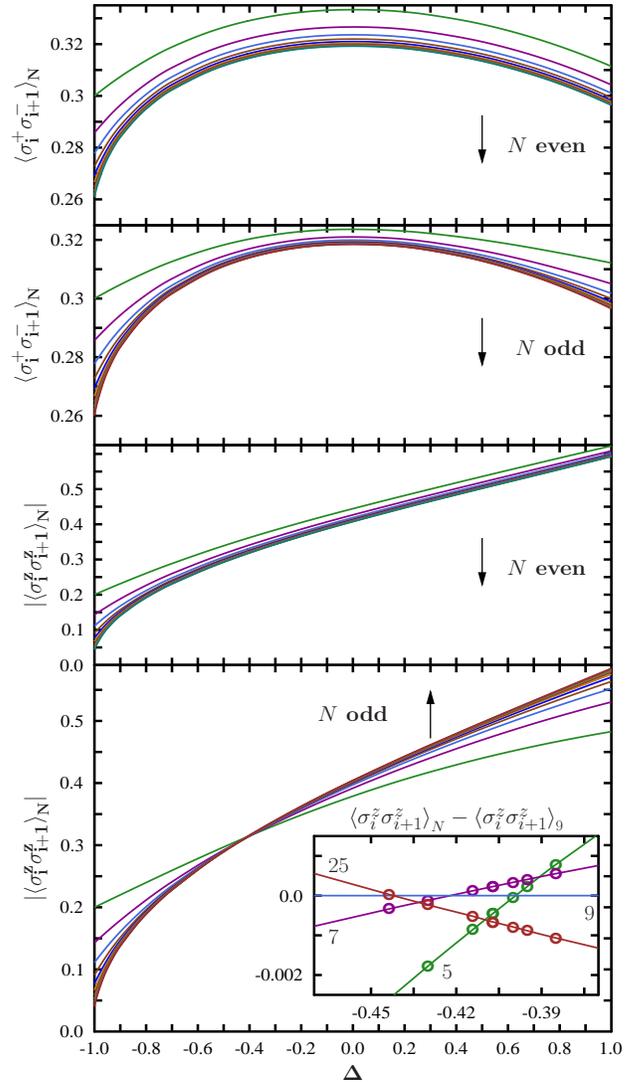}
\caption{Nearest-neighbour correlation functions 
$\langle\sigma_i^+\sigma_{i+1}^-\rangle_{_N}$, 
and $|\langle\sigma_i^z\sigma_{i+1}^z\rangle_{_N}|$ versus $\Delta$ 
for different values of $N$ odd and even. $N$ grows as indicated by the 
arrows, from $5$ to $25$.
The Inset shows the difference 
$\langle\sigma_i^z\sigma_{i+1}^z\rangle_{_N}-
\langle\sigma_i^z\sigma_{i+1}^z\rangle_{_9}$ as a function of $\Delta$, 
for $N$=5, 7, 9, and 25.} 
\label{f.2}
\end{figure}

Before considering the correlation functions, we remind that our results 
are obtained via the numerical diagonalization of the Hamiltonian, so that 
not only some physical observables, but all the coefficients of the ground 
state on the standard basis are available, for $N$ both even and odd.
This allows us to notice the following general features of the ground
state of the XXZ model with $|\Delta|<1$:
\textit{i}) all the elements of the standard basis belonging to the 
sector with $S^z_{tot}=1/2$ for $N$ odd, or $S^z_{tot}=0$ for $N$ even,   
enter the decomposition of the ground state;
\textit{ii}) the coefficients on the standard basis grow with the number of 
antiparallel adjacent spin pairs featuring the corresponding element;
\textit{iii}) keeping the minimum coefficient fixed to unity, all the other
coefficients grow with $\Delta$; 
\textit{iv}) the coefficients are all 
equal in the $\Delta\to -1^+$ limit.

We further underline, due to its relevance for the following discussion,
that the nearest-neighbour correlation functions along the $z$-direction
are negative not only for $\Delta>0$, as expected, but also
for $-1<\Delta\leq 0$. Indeed, although the exchange interaction
in the $z$-direction is ferromagnetic,  configurations where
adjacent spins have opposite components along the $z$-axis are
energetically favoured by  the predominant exchange interaction
in the $xy$-plane, which is 
$(\sigma^x_i\sigma^x_{i+1}+\sigma^y_i\sigma^y_{i+1})$ $=$ 
$2(\sigma^+_{i}\sigma^-_{i+1}+\sigma^-_{i}\sigma^+_{i+1})$.   
For the sake of clarity, in the
lowest panels of Fig.~\ref{f.2} we plot the absolute values of
$\langle\sigma_i^z\sigma_{i+1}^z\rangle_{_N}$.

We now focus upon the finite-size corrections to the quantities
$\langle\sigma_i^+\sigma_{i+1}^-\rangle_{_N}$ and
$\langle\sigma_i^z\sigma_{i+1}^z\rangle_{_N}$: From Fig.~\ref{f.2} it
is clear that $\langle\sigma_i^z\sigma_{i+1}^z\rangle_{_N}$ for odd 
$N$ behaves in a peculiar way: in fact, while 
$\delta_{_N}(\sigma_i^+\sigma_{i+1}^-)$ 
is always positive (the longer the
chain the less correlated adjacent spins are on the $xy$-plane), we find
that for any odd $N$ it exists a value of $\Delta$ where 
$\delta_{_N}(\sigma_i^z\sigma_{i+1}^z)$   
changes its sign, so that, as far as 
the anisotropy is larger than such value, adjacent spins get more and more
correlated along the $z$-direction as the length of the chain grows,
while the opposite occurs otherwise.
It is to be noticed that, at variance with the case of 
$\delta_{_N}({\cal{H}}_{i,i+1})$, the value of $\Delta$ where 
$\delta_{_N}(\sigma_i^z\sigma_{i+1}^z)$ vanishes does 
depend on $N$. This numerical observation is made evident by 
plotting the difference $\langle\sigma_i^z\sigma_{i+1}^z\rangle_{_N}-
\langle\sigma_i^z\sigma_{i+1}^z\rangle_{_{N'}}$ for different values of 
$N$ and fixed $N'$:
should finite-size corrections vanish at a precise value of $\Delta$ 
indipendently on $N$, all the lines in the Inset of Fig.~\ref{f.2}
(where we have chosen $N'=9$) would cross at the same  point, 
which is seen not to be the case. 

Let us now discuss the above results. 
We have found that nearest-neighbour correlation functions along the 
$z$-direction in the 
$S=1/2$ XXZ model with $\Delta>-1$ are always negative due 
to prevailing role of the exchange interaction on the $xy$-plane, which 
evidently favours configurations with antiparallel adjacent spins. 
On the other hand, the necessary occurrence, in the case of $N$ odd, of at 
least one pair of adjacent spins with the same component along the 
$z$-direction, locally frustrates this propensity to antiparallelism.
As such constraint concerns just one spin pair, no matter 
the length of the chain, it may only affect finite-size corrections, 
and its relevance is smeared out as $N$ increases. 
This mechanism of local frustration is responsible for the 
anomalous behaviour of finite-size corrections to 
$\langle\sigma_i^z\sigma_{i+1}^z\rangle_{_N}$ for $N$ odd,
and specifically for the change of sign of 
$\delta_{_N}(\sigma_i^z\sigma_{i+1}^z)$, which ultimately makes it 
possible for $\delta_{_N}({\cal{H}}_{i,i+1})=
-4\delta_{_N}(\sigma_i^+\sigma_{i+1}^-)
+\Delta \delta_{_N}(\sigma_i^z\sigma_{i+1}^z)$ to vanish at 
$\Delta=1/2$, and stay positive for $\Delta>1/2$.
The model whose ground state has the specific structure 
predicted by Razumov and Stroganov is thus found to be a special point
in the finite-$N$ XXZ phase-diagram, as it separates the region where 
adding
spin pairs increases the energy per site from that where the 
longer the chain the lower the energy per site.

From this point of view, the XXZ model in
the $\Delta\to-1^+$ limit is similarly special (for a precise definition
of such model, see Appendix \ref{s.AppendixB}).
Indeed, its ground state
has a peculiar structure:  It is a superposition with identical coefficients of all
the elements of the standard basis within  the $S_{tot}^z=1/2$ ($S_{tot}^z=0$) sector,
for $N$ odd (even). This allows, as shown in Appendix~\ref{s.AppendixB},
for the exact evaluation of the correlation functions. In particular,
similarly to Eq.~\eqref{e.H_N}, one find, for example for $N$ odd,
\begin{eqnarray}\label{e.H_Nsym}
\langle{\cal{H}}_{i,i+1}\rangle_{_N}&=&
-4\langle\sigma^+_i\sigma^-_{i+1}\rangle_{_N}-
\langle\sigma^z_i\sigma^z_{i+1}\rangle_{_N} = \nonumber\\
&=&-1-\frac{1}{N}+\frac{1}{N}=-1~,
 \end{eqnarray} 
and analogously for $N$ even. We thus see that finite-size corrections to 
the  
energy per site (but not to other physical quantities) do vanish for this model as well.

We notice that there exists another class of models whose expression for
the energy per site is given by Eq.~(\ref{e.E-special}), namely any
antiferromagnetic XYZ Heisenberg system on a bipartite lattice with
periodic boundary conditions, in the presence of a uniform magnetic field
whose value fulfills the condition given by Kurman et al. in
Ref.~\cite{KTM-82}.
When such condition is fullfilled, the
ground state factorizes, meaning that it gets the unexpected
classical-like structure of a tensor product of single-spin states. This
phenomenon has recently attracted much attention, due to its relevance as
far as the entanglement properties are concerned
\cite{RoscildeEtal04,RoscildeEtal05,GiampaoloEtal08,GiampaoloAI08}.
In this respect, one should notice that while
the absence of finite-size corrections in the models studied by Kurmann et
al., as well as in the $\Delta=-1$ ferromagnetic isotropic model, directly 
follows from the factorized structure of the ground state,
thus characterizing the expectation value of whatever physical
observable, in the Razumov-Stroganov state, as well as in the  
$\Delta\to-1^+$ ground state,  finite-size corrections vanish
only in the energy per site, following a precise cancellation in the
expectation value of the Hamiltonian. 


\section{Entanglement properties}\label{s.entanglement}

The investigation of entanglement properties has revealed a powerful tool 
in 
studying spin models, particularly in unveiling peculiar properties of 
the ground state~\cite{ArnesenBV01,OsborneN02,WangZ02,S-03,VerstraeteEtal04,OsterlohEtal02,
AmicoEtal08}, as in the case of the XYZ 
antiferromagnet in the uniform field mentioned above~\cite{RoscildeEtal04}.
Several authors have specifically addressed the analysis of the XXZ 
model in terms of entanglement properties 
\cite{VidalEtal03,Korepin04,JinK04,GuLL03,GuTL05,MeyerEtal04}, and 
more recently the case of $\Delta=1/2$ and $N$ odd has been 
studied in such context \cite{CCN-09}.
In this framework, the idea that the peculiar structure of the Razumov-Stroganov state 
might be related with properties of some entanglement measure is 
suggestive. We have therefore developed the same type of analysis 
presented in Section~\ref{s.other_models}, for the bipartite entanglement 
between two $S=1/2$ spins separated by $r$ lattice spacings, as measured by the 
concurrence~\cite{HillW97,Wootters98,AmicoEtal04} 
\begin{equation}
C_{r,N}=2\max\{0,C'_{r,N}\}~,
\label{e.C}
\end{equation}
where
\begin{equation}
C'_{r,N}=\frac{1}{2}\left|\langle\sigma_i^x\sigma_{i+r}^x\rangle_{_N}\right|
	-\frac{1}{4}\sqrt{\left(1+
	\langle\sigma_i^z\sigma_{i+r}^z\rangle_{_N} \right)^2
	- 4\langle\sigma_i^z\rangle_{_N}^2 }~.
\label{e.C'}
\end{equation}
\begin{figure}
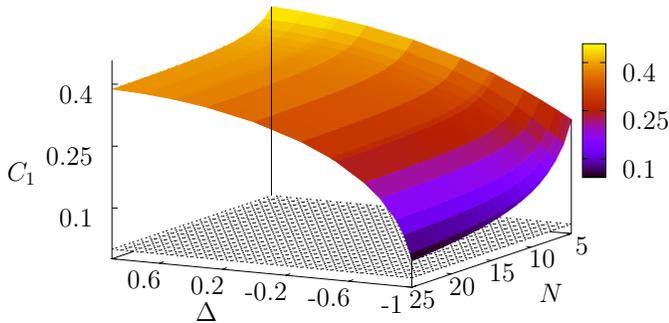

\centering\scalebox{0.7}{\makebox[\textwidth][l]{\input conc1.tex}}
\caption{$C_{1}$ versus $\Delta$ and $N$}
\label{f.3}
\end{figure}

We have numerically evaluated $C_{r,N}$ for various $\Delta$, $r$, and
$N$.
In Figs.~\ref{f.3} and \ref{f.4} we show $C_{1,N}$ and $C_{2,N}$ versus
 $N$ and $\Delta$. Apart from very small oscillations between odd
and even $N$, essentially due to different finite-size corrections in the 
correlation functions
and to the fact that $\langle\sigma_i^z\rangle_{_N}^2$ in Eq.~(\ref{e.C'})
is positive for $N$ odd and vanishes for $N$ even, no peculiar behaviour
is observed. 
The concurrence between nearest neighbours $C_{1,N}$ is always finite 
and is larger for larger $\Delta$, i.e. for more
marked antiferromagnetic exchange along the $z$-direction (as expected
after the analysis presented in Ref.~\cite{FubiniEtal06,AmicoEtal06}).
All pairs of non-adjacent spins are disentangled in the Razumov-Stroganov state, no 
matter the value of $N$, but this is not a specific feature 
of the model (\ref{e.HRS}), as
already shown by the overall picture presented in Ref.~\cite{S-03}.
After all, and despite the particular structure of its ground 
state, the case  $\Delta=1/2$ and $N$ odd is not found
to display peculiar features, at least as far as pairwise entanglement is
concerned. The way transverse and longitudinal correlation functions
combine in Eq.~(\ref{e.C'}) does not cause any relevant change in the 
finite-size corrections of the concurrence.

\begin{figure}
\centering\scalebox{0.7}{\makebox[\textwidth][l]{\input conc2.tex}}
\caption{$C_{2}$ versus $\Delta$ and $N$}
\label{f.4}
\end{figure}

\begin{figure}
\includegraphics[width=1\linewidth]{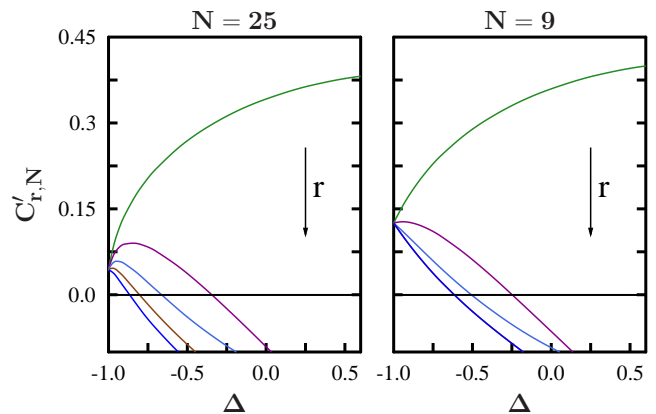}
\caption{$C'_{r,N}$ as a function of $\Delta$ for $N=25$ (left) and 
$N=9$ (right) for $r\le 5$.
}
\label{f.5}
\end{figure}

Let us now consider the other point where finite-size corrections to the 
energy per site vanish, i.e. the ferromagnetic isotropic point 
$\Delta=-1$. As a matter of fact, the model
in  the $\Delta\to -1^+$ limit displays features which 
deserve some comments. In the thermodynamic limit, all the $C_r$ are 
expected~\cite{AmicoEtal06} to switch on as $\Delta\to -1^+$, which 
implies, due to the 
monogamy inequality \cite{CoffmanKW00,OsborneV05}, that all the 
concurrences 
vanish at the ferromagnetic 
isotropic point $\Delta=-1$, consistently with the fact that the 
ground state of the XXZ chain with $\Delta\le -1$ is fully separable. On the other
hand, when $N$ is finite (no matter if even or odd), not only all the $C_{r,_N}$ 
switch on as   $\Delta\to -1^+$, but they all flow to the same value,
as seen in Fig.~\ref{f.5}.
In particular, from the exact evaluation of the correlation functions
reported in Appendix~\ref{s.AppendixB}, 
one can find that
\begin{equation}
\lim_{\Delta\to -1^{^+}} C_{r,N} = 
\left\{\begin{array}{ll}
\frac{N+1}{2N}\left(1-\sqrt{\frac{N-3}{N+1}}\right)~~&\forall r \mbox{ , 
$N$ odd}~;\\
\frac{1}{N-1}&\forall r \mbox{ , $N$ even}~.
\end{array}
\right.
\label{e.CrNisoferro}
\end{equation}
Moreover, when $\Delta< -1$ the result $C_{r,N}=0$ for all $r$ stays valid also for 
finite $N$; therefore, and at variance with the thermodynamic limit,
a discontinuity of all the concurrences occurs at $\Delta=-1$.

We underline that the one-tangle~\cite{CoffmanKW00} 
$\tau=1-\sum_\alpha\langle\sigma^\alpha\rangle_{_N}^2$ in the XXZ model with 
periodic boundary conditions and $|\Delta|<1$ is different from zero and does not 
depend on $\Delta$ (being $\tau=1-1/N^2$ for $N$ odd, and  $\tau=1$ for $N$ even),
while for $\Delta\leq-1$, $\tau$ vanishes.
The observed discontinuity at $\Delta=-1$ (at variance with the one characterizing
the concurrence) is not related with the finite size of the system, 
as it obviously survives in the thermodynamic limit.
It is rather connected with the onset of critical behaviour for $|\Delta|<1$,
which implies the vanishing of the magnetizations
and a consequently maximal entanglement content.

We understand the mechanism leading to such discontinuities as 
follows: From our results we observe, as stated in 
Section~\ref{s.other_models}, that the ground state of the XXZ 
model for $\Delta\to -1^+$ may be 
written as $P^+\Pi_i|\varphi\rangle_i$ where  $P^+$ is the 
projector over the Hilbert subspace of the chain corresponding to 
$S^z_{tot}=1/2$, and $|\varphi\rangle$ 
is a single-spin pure state which may be chosen at will, provided it is 
not one of the two eigenvectors of $\sigma^z$ (so as to ensure the 
projection does not  vanish).
The notation indicates that all the spins 
along the chain are in the same single-spin pure state: In fact, the fully 
separable state $\Pi_i|\varphi\rangle_i$ is just one of the infinite 
ground states of the ferromagnetic XXX chain. 
We therefore see that it is the projection over the
$S^z_{tot}=1/2$ Hilbert subspace which injects
entanglement into the fully separable ground state of the ferromagnetic XXX
model. We notice that such projection is not a local operation, and it  
can hence modify the entanglement content of the system.
The above reasoning is straightforwardly extended to the $N$ even case, 
replacing $P^+$ with $P^0$, i.e. with 
the projector over the $S^z_{tot}=0$ Hilbert subspace.
In other terms, the easy-plane character of 
the XXZ model with $|\Delta|< 1$, which is embodied in the condition 
$S^z_{tot}= 1/2$ for $N$ odd, or $S^z_{tot}=0$ for $N$ even, causes the 
ground state to develop long-ranged 
pairwise entanglement as the ferromagnetic isotropic point is approached 
from above.

Since pairwise entanglement between two spins does not exhaust 
the possible types of entanglement,  and there might
be other measures that exhibit a special behaviour near or at 
$\Delta=1/2$, we have numerically evaluated the entanglement entropy for a 
wide range
of bipartition of the full chain, and for various $N$ and $\Delta$.
In the particular case where the bipartition consists of two blocks
of adjacent spins, we find a very good agreement with the theoretical 
predictions  of conformal field theory~\cite{CC-04}. Such agreement
is observed not only for $\Delta = 1/2$ (as already emphasized in 
Ref.~\cite{CCN-09}) but
all over the interval  $-1<\Delta<1$, and $N$ both even and odd.
It is worth underlining that such agreement is observed to hold  
for blocks of at least two spins with great accuracy, even for rather short chains, 
with $N$ as small as $5$.


\section{Conclusions}\label{s.conclusions}

Our study of the ground state structure as related with finite-size 
effects in the one-dimensional $S=1/2$ XXZ model 
with periodic boundary conditions in its critical phase 
($T=0$ and $|\Delta|<1$ )
has highlighted some relevant features; these particularly emerge for 
$\Delta=1/2$ and $N$ odd, as well as in the ferromagnetic isotropic limit, 
$\Delta\to -1^+$ (for $N$ both even and odd). Indeed,  
these two models share  a common property, namely the fact that finite-size 
corrections to the energy per site vanish, despite being finite as far as 
other physical quantities are concerned.

In the former case, we have found that the model represents a special
point in the $N$-odd sector (where its ground state has the structure of
the Razumov-Stroganov state), as it separates the region where adding spin
pairs increases the energy per site of the chain, from that where the
reverse holds.  We think this result, besides giving some further insight
into the behaviour of the magnetic XXZ model, could get a specific meaning
for the fermionic models whose Hamiltonian is mapped into the $S=1/2$ XXZ
one (see for example Ref.~\cite{FNS-03}), possibily in terms of a
physical mechanism related with a properly defined chemical potential.

In the $\Delta\to -1^+$ limit, the peculiarities of 
the model do also arise from the very specific
(despite not as intriguing as that of the Razumov-Stroganov state) 
structure of the ground  state, 
which guarantees the exact cancellation of finite-size corrections 
to the energy per site, for $N$ both even and odd.
However, in this case the peculiar ground state structure most 
relevantly reflects on bipartite entanglement: all the 
concurrences are finite and get the same value, no matter the 
distance between the two spins considered. 
In fact, we have found that there is a region, while approaching the
ferromagnetic isotropic point from above, where any two spins along the
chain are entangled, given that the finite size of the chain prevents the
{\it monogamy of the entanglement} to force the vanishing of all the
concurrences; in such region the entanglement between a selected
spin pair can be varied not only by tuning $\Delta$ but also by properly
choosing $N$, a possibility which might be relevant in the design of 
quantum devices.

Finally, we recall that the two above mentioned models, besides sharing a 
somewhat specific finite-size behaviour, present another rather 
exceptional peculiarity, namely the possibility of accessing the exact 
structure of their ground state. Indeed, in both cases, the (suitably 
normalized) coefficients of the ground state in the standard basis are all 
integers:  This naturally sets the models into the realm of 
combinatorics, and has permitted us to derive new exact analytic 
expressions for their correlation functions.

\begin{acknowledgments}

We thank Miguel Iba\~{n}ez de Berganza, Alessandro Cuccoli, Andrei Pronko and Ruggero Vaia 
for useful discussions.
FC acknowledges partial support from the European Science Foundation program
INSTANS, and from MIUR, under the PRIN grant 2007JHLPEZ.
\end{acknowledgments}


\appendix
\section{Some exact results at $\Delta=1/2$}\label{s.AppendixA}

We report here the exact values of the second neighbour longitudinal 
two-point correlation function $\langle\sigma^z_i\sigma^z_{i+2}\rangle_N$, 
for $N=3,5,\dots, 15$:
\begin{eqnarray}
\langle \sigma^z_{i}\sigma^z_{i+2}\rangle_{3}&=&-\frac{1}{3}\nonumber\\
\langle \sigma^z_{i}\sigma^z_{i+2}\rangle_{5}&=&\frac{1}{25}\nonumber\\
\langle \sigma^z_{i}\sigma^z_{i+2}\rangle_{7}&=&\frac{4}{49}\nonumber\\
\langle \sigma^z_{i}\sigma^z_{i+2}\rangle_{9}&=&\frac{28}{297}\\
\langle \sigma^z_{i}\sigma^z_{i+2}\rangle_{11}&=&\frac{157}{1573}\nonumber\\
\langle \sigma^z_{i}\sigma^z_{i+2}\rangle_{13}&=&\frac{191}{1859}\nonumber\\
\langle \sigma^z_{i}\sigma^z_{i+2}\rangle_{15}&=&\frac{1732}{16575}\nonumber
\end{eqnarray}
Note that  all these values are rational numbers. This remarkable
property allows for their infinite-precision numerical 
evaluation, and permits  to derive the exact expressions 
of the correlation functions for arbitrary (odd) values of $N$. We 
report below our results for the  $r^{\textrm{th}}$-neighbour
longitudinal and transverse two-point correlation functions, $r=2,3,4,5$:
\begin{equation}\label{e.app.gzzj=2}
\langle\sigma_i^z\sigma_{i+2}^z\rangle_{_N} = 
\frac{7}{2^6}-\frac{3}{2^6} \left(\frac{227+22 N^2}{N^2 (N^2-4)}\right)
\end{equation}
\begin{equation}
\langle\sigma_i^+\sigma_{i+2}^-\rangle_{_N} = 
\frac{41}{2^8} + \frac{105}{2^8} \left(
\frac{1+2 N^2}{N^2 \left(N^2-4\right)}\right)
\end{equation}
\begin{widetext}
\begin{equation}
\langle\sigma_i^z\sigma_{i+3}^z\rangle_{_N}  = 
-\frac{401}{2^{12}} +\frac{45}{2^{12}}\left(
\frac{-2205+5324 N^2+26 N^4+212 N^6}{N^4 (N^2-4)^2}\right)
\end{equation}
\begin{equation}
\langle\sigma_i^+\sigma_{i+3}^-\rangle_{_N} = 
\frac{4399}{2^{15}} + \frac{9}{2^{15}} \left(
\frac{-11025+1276 N^2-7358 N^4+4516 N^6}{N^4 \left(N^2 - 4\right)^2}\right)
\end{equation}
\begin{multline}
\langle\sigma_i^z\sigma_{i+4}^z\rangle_{_N} = 
\frac{184453}{2^{22}} - \frac{3}{2^{22}
N^4 \left(N^2-4\right)^3  \left(N^2-16\right)}
\Big(16094658825  -4071808726 N^2\\
+1675416103 N^4-269157300 N^6
+ 41497687 N^8+2281766 N^{10}\Big)
\end{multline}
\begin{multline}
\langle\sigma_i^+\sigma_{i+4}^-\rangle_{_N} = 
\frac{1751531}{2^{24}} + \frac{3}{2^{24}
N^4  \left(N^2-4\right)^3 \left(N^2-16\right)} \Big(
4695690825 + 2413434266 N^2 \\
- 276248393 N^4+556663980 N^6 
-134362697 N^8 + 10926614 N^{10} \Big)
\end{multline}
\begin{multline}
\langle\sigma_i^z\sigma_{i+5}^z\rangle_{_N} = 
-\frac{95214949}{2^{31}}+
\frac{3}{2^{31}
N^6 \left(N^2-4\right)^4\left(N^2-16\right)^2} \Big(
3019990307709375 -1994398549102575 N^2 \\
+ 22018044125468N^4 -149133792747084 N^6
+ 28928020131522 N^8  -4476210910162 N^{10} \\
+ 693645483372 N^{12} -43191843804N^{14}
+ 2051652263 N^{16}
\Big)
\end{multline}
\begin{multline}
\langle\sigma_i^+\sigma_{i+5}^-\rangle_{_N} = 
\frac{3213760345}{2^{35}}+\frac{3}{2^{35}
N^6 \left(N^2-4\right)^4\left(N^2-16\right)^2} \Big(
4227986430793125  -565146993503925 N^2 \\
-383849970492812 N^4 -183487079646084 N^6 + 70690548132342 N^8  -35899621486502 N^{10}\\
+ 8073828981732 N^{12} -796229443764 N^{14}+ 28577677213 N^{16}
\Big)
\end{multline}
\end{widetext}


\section{The $\Delta\to-1^+$ model}\label{s.AppendixB} 

We provide here some analytic results on the XXZ chain of length $N$,
in the  $\Delta\to-1^+$ limit. It is worth emphasizing that,
by construction,  this model differs from the isotropic 
$\Delta=-1$ ferromagnet. It  is defined as a limit from the critical region, 
and its ground state, together with the expectation value of all physical quantities,
is also to be intended as the result of this limit. In particular,  
the model is critical, with 
$S^z_{tot}=1/2$ for $N$ odd, and  $S^z_{tot}=0$ for $N$ even.
Setting $\Delta=-1+\epsilon$, there obviously are finite corrections  
to all the exact expressions we provide below, but they all vanish as $\epsilon\to 0$.

It is easily seen, from analytic considerations, and numerical
experiments, that the ground state of the XXZ chain in the $\Delta\to-1^+$ limit is
a superposition with identical coefficients (which we set to unity)
of all the elements of the  standard basis corresponding to 
$S^z_{tot}=1/2$ for $N=2n+1$ (odd), and to 
$S^z_{tot}=0$ for $N=2n$ (even). 
Such elements are herafter indicated by $|e_l\rangle$.
Their number is $\left(^{N}_{\,n}\right)$;
the norm of the state consequently reads ${\cal{N}}=\sqrt{\left(^{N}_{\,n}\right)}$.
Due to the peculiar structure of the ground state, 
the evaluation of correlation functions reduces to a simple combinatorial 
enumeration, as shown below.

Let us consider the correlation functions on the $xy$-plane: it is
$\langle\sigma^+_i\sigma^-_j\rangle_{_N}={\cal{N}}^{-2}\sum_{lm}\langle 
e_l|\sigma^+_i\sigma^-_j|e_m\rangle$, and the elements $|e_m\rangle$ that 
do contribute to the double sum 
are just those with $\sigma^z_i|e_m\rangle=- |e_m\rangle$, and 
$\sigma^z_j|e_m\rangle= |e_m\rangle$; each of them will contribute exactly $1$
and their number is $\left(^{N-2}_{\,n-1}\right)$, so that
\begin{eqnarray}
\langle\sigma^+_i\sigma^-_j\rangle_{_N}&=&\frac{N+1}{4N}\,,\qquad\quad N\ \mathrm{odd,}\\
\langle\sigma^+_i\sigma^-_j\rangle_{_N}&=&\frac{N}{4(N-1)}\,,\qquad N\ \mathrm{even,} 
\end{eqnarray}
no matter the distance between site $i$ and $j$. 

As for the correlation functions along the $z$-axis, it is
$\langle\sigma^z_i\sigma^z_j\rangle={\cal{N}}^{-2}\sum_{lm}\langle 
e_l|\sigma^z_i\sigma^z_j|e_m\rangle$, and each term 
in the sum is easily seen to contributes with $\pm \delta_{lm}$. 
The various terms  can be collected, according to the 
orientation of the $i^{\rm th}$ and $j^{\rm th}$ spins in each $|e_m\rangle$, into 
four groups, so as to obtain
\begin{eqnarray}
\langle\sigma^z_i\sigma^z_j\rangle_{_N}&=&{\cal{N}}^{-2} \left[
\left(^{N-2}_{\quad n}\right)
-\left(^{N-2}_{\ n-1}\right)
-\left(^{N-2}_{\ n-1}\right)
+\left(^{N-2}_{\ n-2}\right)
\right]\nonumber\\
&=&\frac{(N-2n)^2 -N}{N(N-1)}\,,
\end{eqnarray}
and hence
\begin{eqnarray}
\langle\sigma^z_i\sigma^z_j\rangle_{_N}&=&-\frac{1}{N}\,,\qquad\quad N\ \mathrm{odd,}\\
\langle\sigma^z_i\sigma^z_j\rangle_{_N}&=&-\frac{1}{N-1}\,,\qquad N\ \mathrm{even,} 
\end{eqnarray} 
no matter the distance between site $i$ and 
$j$.

The fact that two-point correlation functions do not depend on the 
distance is a direct consequence of the particular structure of the ground 
state, whose non vanishing coefficients on the standard basis are all 
identical. It is worth mentioning that each of these correlation functions 
assumes the same value on the chain of length $2n-1$ and $2n$. Finally, 
notice that these correlation functions have different values with respect 
to the $\Delta=-1$ isotropic ferromagnetic model. The consequent 
discontinuity vanishes in the thermodynamic limit.



\begin{thebibliography}{99}

\bibitem{DiVincenzoEtal00}
D.P.~DiVincenzo, D.~Bacon, J.~Kempe, G.~Burkard, and K.B.~Whaley,
Nature 408, 339 (2000);
\bibitem{MeierLL03}
F.~Meier, J.~Levy, and D.~Loss, Phys. Rev. B \textbf{68}, 134417 (2003);
\bibitem{AAVVnato06}
For a general review see for instance
{\em Quantum information processing: from theory to experiment},
vol.199 NATO Scienze Series: Computer and Systems Sciences, Eds.
D.G.~Angelakis, M.~Christandl, A.~Ekert, A.~Kay and S.~Kulik,
(2006);

\bibitem{Sachdev99}
S.~Sachdev, {\em Quantum Phase Transitions} (Cambridge University Press,
Cambridge 1999);

\bibitem{YangY66}
C.N.~Yang and C.P.~Yang., Phys. Rev. {\bf 147}, 303 (1966);
\bibitem{KatsuraS70}
S.~Katsura and M.~Suzuki, J. Phys. Soc. Japan {\bf 28}, 255 (1970);
\bibitem{Minami04}
K.~Minami, J. Mag. Mag. Mater. {\bf 270}, 104 (2004);

\bibitem{KTM-82}
J.~Kurmann, H. Thomas and G. M\"uller,
Physica A {\bf 112}, 235 (1982);
\bibitem{RoscildeEtal04}
T. Roscilde, P. Verrucchi, A. Fubini, S.~Haas, and V.~Tognetti,
Phys. Rev. Lett. {\bf 93}, 167203 (2004);

\bibitem{ABB-88} F.C.~Alcaraz, M.N.~Barber and M.T.~Batchelor, 
Ann.~Phys.~N.Y. \textbf{182}, 280 (1988);
\bibitem{S-01} Yu.~Stroganov,
J.~Phys.~A:~Math.~Gen. \textbf{34} ,L179 (2001);
\bibitem{RS-01} A.V.~Razumov and Yu.G.~Stroganov,
J.~Phys.~A:~Math.~Gen. \textbf{34}, 3185  (2001);


\bibitem{PRdGN-02} P.A.~Pearce, V.~Rittenberg, J.~de~Gier and B.~Nienhuis,
J.~Phys.~A:~Math.~Gen. \textbf{35}, L661 (2002); 
\bibitem{FNS-03}
P.~Fendley, B.~Nienhuis and K.~Schoutens,
J.~Phys.~A:~Math.~Gen. \textbf{36} 12399 (2003);
\bibitem{DFZJ-05} P.~Di~Francesco and P.~Zinn-Justin,
J.~Phys.~A:~Math.~Gen. \textbf{38}, L815 (2005); 
\bibitem{dG-05} J.~de~Gier,
Discr.~Math. \textbf{298}, 365  (2005);

\bibitem{RS-01b}A.V.~Razumov and Yu.G.~Stroganov,
J.~Phys.~A:~Math.~Gen. \textbf{34},  5335  (2001);
\bibitem{BdGN-01} M.T.~Batchelor, J.~de Gier and B.~Nienhuis, 
J.~Phys.~A:~Math.~Gen. \textbf{34}, L265 (2001);

\bibitem{B-08} L. Banchi, `Struttura e propriet\`a di entanglement
dello stato fondamentale nella catena di spin di Razumov-Stroganov',
Master thesis (in italian), September 2009, available at:\par
\texttt{http://theory.fi.infn.it/colomo/tesibanchi.pdf};
\bibitem{CCN-09} B.~Nienhuis, M.~Campostrini, P.~Calabrese, 
J.~Stat.~Mech. P02063 (2009);

\bibitem{KMST-05} N.~Kitanine, J.M.~Maillet, N.A.~Slavnov, and V.~Terras,
J.~Stat.~Mech. 0509-L002 (2005);
\bibitem{SS-07} J.~Sato and M.~Shiroishi,
J.~Phys.~A:~Math.~Theor. \textbf{40},  8739 (2007);

\bibitem{B-71} R. Baxter, 
Phys.~Rev.~Lett. \textbf{26}, 834 (1971);

\bibitem{RoscildeEtal05}
T.~Roscilde, A.~Fubini, P.~Verrucchi, S.~Haas, and V.~Tognetti, 
Phys. Rev. Lett. {\bf 94}, 147208 (2005);
\bibitem{GiampaoloEtal08}
S.M.~Giampaolo, F.~Illuminati, P.~Verrucchi, and S.~De Siena,
Phys.~Rev.~ {\bf A 77},  012319 (2008);
\bibitem{GiampaoloAI08}
S.M.~Giampaolo, G.~Adesso, and F.~Illuminati, Phys. Rev. Lett. {\bf 100}, 
197201 (2008);

\bibitem{ArnesenBV01}
M.C.~Arnesen, S.~Bose, and  V.~Vedral, Phys. Rev. Lett. {\bf 87},
017901 (2001);
\bibitem{OsborneN02}
T.J.~Osborne and  M.A.~Nielsen, Phys. Rev. A {\bf 66}, 032110
(2002);
\bibitem{WangZ02}
X.~Wang and P.~Zanardi, Phys. Lett. A {\bf 301}, 1 (2002);
\bibitem{OsterlohEtal02}
A.~Osterloh, L.~Amico, G.~Falci, and R.~Fazio, Nature (London)
{\bf 416}, 608 (2002);
\bibitem{S-03}
O.F.~Sylju\aa sen, Phys. Rev. A {\bf 68}, 060301 (2003); {\em ibid.}
Phys.~Lett.~A {\bf 322}, 25 (2004);
\bibitem{VerstraeteEtal04}
F. Verstraete, M. Popp, and J. I. Cirac, Phys. Rev. Lett. {\bf
92}, 027901 (2004);
\bibitem{AmicoEtal08}
L.~Amico, R.~Fazio, A.~Osterloh, and V.~Vedral, Rev. Mod. Phys. {\bf 80}, 
517-576 (2008);

\bibitem{VidalEtal03}
G. Vidal, J. I. Latorre, E. Rico, and A. Kitaev, 
Phys. Rev. Lett. {\bf 90}, 227902 (2003);
\bibitem{Korepin04}
V.E.~Korepin, Phys.~Rev.~Lett. {\bf 92}, 096402 (2004);
\bibitem{JinK04} 
B.-Q.~Jin and V.E.~Korepin, Phys. Rev. A, {\bf 69}, 062314 (2004);
\bibitem{GuLL03}
S-J.~Gu, H-Q.~Lin, and Y-Q.~Li, 
Phys. Rev. A {\bf 68} 042330 (2003);
\bibitem{GuTL05}
S-J.~Gu, G.-S.~Tian, and H.-Q.~Lin,
Phys.~Rev.~A {\bf 71}, 0-52322 (2005);

\bibitem{MeyerEtal04}
T.~Meyer, U.V.~Poulsen, K.~Eckert, M.~Lewenstein, and D.~Bruss,
Int. J. Quant. Inf. {\bf 2}, 149 (2004);


\bibitem{HillW97}
S. Hill and W.K.~Wootters, Phys. Rev. Lett. {\bf 78}, 5022 (1997);
\bibitem{Wootters98}
W.K.~Wootters, Phys. Rev. Lett. {\bf 80}, 2245 (1998);
\bibitem{AmicoEtal04}
 L.~Amico, A. Osterloh, F.~Plastina, R.~Fazio, and G.~M.~Palma,
Phys. Rev. A {\bf 69}, 022304 (2004);


\bibitem{FubiniEtal06}
A.~Fubini, T.~Roscilde, M.~Tusa, V.~Tognetti, and P.~Verrucchi,
Eur.~Phys.~J.~D {\bf 38}, 563 (2006);

\bibitem{AmicoEtal06}
L. Amico, F. Baroni, A. Fubini, D. Patan\`e, V. Tognetti, and P.
Verrucchi, Phys. Rev. A {\bf 74}, 022322 (2006);

\bibitem{CoffmanKW00}
V.~Coffman, J.~ Kundu, and W.K.~ Wotters,
Phys. Rev. A {\bf 61} 052306 (2000);

\bibitem{OsborneV05}
T.J.~Osborne and F.~Verstraete, Phys. Rev. Lett. {\bf 96}, 220503 (2006);

\bibitem{CC-04} Pasquale Calabrese and John Cardy,
J.~Stat.~Mech.  P06002 (2004).

\end{thebibliography}
\end{document}